\documentclass[aps,prx,reprint,superscriptaddress,showpacs]{revtex4-1}

\usepackage{amsmath}    
\usepackage{graphicx}   
\usepackage{color}      
\usepackage{subfigure}  
\usepackage{hyperref}   
\usepackage{todonotes}
\usepackage{nicefrac}
\usepackage[utf8]{inputenc}

\begin{document}

\title{Evidence for an orbital dependent Mott transition in the ladders of (La,\,Ca)$_x$Sr$_{14-x}$Cu$_{24}$O$_{41}$ derived by  electron energy-loss spectroscopy.}
\author{Friedrich Roth}
\affiliation{Institute of Experimental Physics, TU Bergakademie Freiberg, Leipziger Stra\ss e 23, D-09599 Freiberg, Germany}
\author{Alexandre Revcolevschi}
\affiliation{Laboratoire de Physico-Chimie de l’État Solide, Université Paris-Sud, 91405 Orsay, France}
\author{Bernd B\"uchner}
\affiliation{IFW Dresden, P.O. Box 270116, D-01171 Dresden, Germany}
\author{Martin Knupfer}
\affiliation{IFW Dresden, P.O. Box 270116, D-01171 Dresden, Germany}
\author{J\"org Fink}
\affiliation{IFW Dresden, P.O. Box 270116, D-01171 Dresden, Germany}
\affiliation{Max Planck Institute for Chemical Physics of Solids, D-01187 Dresden, Germany}
\affiliation{Institute for Solid-State and Material Physics,  Technical University  Dresden, D-01062 Dresden, Germany}
\date{\today}

\begin{abstract}
The knowledge of the charge carrier distribution among the different orbitals of Cu and O is a precondition for the understanding of the physical properties of various Cu-O frameworks. We employ electron energy-loss spectroscopy to elucidate the charge carrier plasmon dispersion in (La, Ca)$_x$Sr$_{14-x}$Cu$_{24}$O$_{41}$ in dependency of $x$ as well as temperature. We observe that the energy of the plasmon increases upon increasing Ca content, which signals an internal charge redistribution between the two Cu-O subsystems. Moreover, contrary to an uncorrelated model we come to the conclusion that the holes transferred to the Cu$_2$O$_3$ ladders are mainly located in the bonding and not in the anti-bonding band. This is caused by an orbital dependent Mott transition.
\end{abstract}

 \maketitle

\section{Introduction}

Thirty-three years after the discovery of high-temperature superconductivity in cuprates, \cite{Bednorz1986} there is still no generally accepted theory of the mechanism of this phenomenon. The reason for this is that we do  not understand the normal state strongly correlated electronic structure of CuO$_2$ layers doped with holes. A starting model for the electronic structure of cuprates was the one-band Hubbard or $t-J$ model~\cite{Zhang1988}. In this model, a hole in the divalent Cu and a hole in the surrounding square  of four O ions form a singlet that describes the electronic structure in a one-band Hamiltonian. In this way, internal degrees of freedom in the Cu-O square are not taken into account. This model replaced the much  more complicated three-band models~\cite{Emery1987} of these charge transfer insulators~\cite{Zaanen1985}. There is an ongoing debate whether the $t-J$ model describes the measured band dispersion and whether it is able to explain high-$T_c$ superconductivity~\cite{Lau2011,Ebrahimnejad2014,Ebrahimnejad2016}.

\par

A way to test the general validity of the $t-J$ model is to look at  compounds  which contain  Cu-O structures that are similar to the two-dimensional CuO$_2$ layers in high-$T_c$ superconductors. In a recent study of T-CuO~\cite{Adolphs2016}, where the $C_4$ rotational symmetry is broken, the authors came to the conclusion that there occur strong deviations  from the  Zhang-Rice singlet picture.
 
\par

Another possibility to study the validity of the Zhang-Rice singlet picture is to investigate the electronic structure of cuprates with a reduced dimension, e.\,g., the quasi-one-dimensional compound (La,\,Y,\,Sr,\,Ca)$_{14}$Cu$_{24}$O$_{41}$~\cite{Vuletic2006}. These materials are composed of alternating stacks of edge-sharing CuO$_2$ chains and two-leg Cu$_2$O$_3$ ladders.  These are separated by strings of Sr, Ca, and La atoms. The subsystems are arranged in layers, and the layers are oriented in the crystallographic $a,c$ plane, while they are stacked in an alternating manner along the perpendicular $b$ axis~\cite{McCarron1988,Siegrist1988}. Electronically, Ca$_{x}$Sr$_{14-x}$Cu$_{24}$O$_{41}$ is inherently doped with six holes per formula unit. Despite this doping level, the parent compound Sr$_{14}$Cu$_{24}$O$_{41}$ shows semiconducting behavior. The substitution of Sr by Ca causes the materials to become conductive, and at a high substitution level ($x$\,=\,13.6) even superconductivity under high pressure has been observed~\cite{Uehara1996}, which renders Ca$_{13.6}$Sr$_{0.4}$Cu$_{24}$O$_{41}$ the first superconducting copper oxide material with a non-square lattice. Furthermore, charge order has been reported and discussed for Ca$_{x}$Sr$_{14-x}$Cu$_{24}$O$_{41}$, indicating a competition between superconductivity and charge ordering in the ladder sub-system. As the substitution of Sr by Ca is iso-electronic, the electronic changes in the materials have been explained by the occurrence of chemical pressure and thereby induced (i) charge transfer from the chains to the ladders, and (ii) singlets on the rungs of the ladders~\cite{Vuletic2006}. The hole distribution in Ca$_{x}$Sr$_{14-x}$Cu$_{24}$O$_{41}$ as a function of $x$ has been controversially discussed, the number of holes in the ladders scatters by almost 100\%~\cite{Kato1996,Osafune1997,Magishi1998,Nuecker2000,Isobe2000,Rusydi2007,Deng2011,Ilakovac2012,Huang2013,Bugnet2016}. 

\par

Moreover, there is a strong spread in the distribution of the holes on the two O atoms, two on the legs and one on the rung  in the  Cu$_2$O$_3$ units of the ladders. If one assumes an equal distribution of the holes in the square of O atoms surrounding the Cu sites, as in the $t-J$ model, one would expect the same number of holes in the rungs and in the legs. Most of the previous studies came to the conclusion that the holes are  predominantly on the rungs~\cite{Vuletic2006}. On the other hand, a more recent study concluded that for the holes accommodated on the ladders, leg sites are preferred  to rung sites~\cite{Ilakovac2012}. 

\par

The two-leg ladders are a representative example between one-dimensional electronic systems forming a Luttinger liquid and  the two-dimensional Cu-O layers in high-$T_c$ superconductors. DFT calculations came to the conclusion that the low energy physics of the ladders can be described by two bands close to the Fermi level: a bonding and an anti-bonding band along the $c$ direction~\cite{Arai1997,Mueller1998}. Part of these bands has been detected in ARPES studies~\cite{Koitzsch2010}. For the interpretation of magnetic and interband charge excitations, it was necessary to take into account correlation effects which were treated in the $t-J$ approximation~\cite{Kumar2019}. Certainly, it would be interesting to perform experiments on these systems to test whether  one has to go beyond this approximation.

\par

The influence of electronic correlations in this class of materials is profound. The interplay of spin and charge degrees of freedom in both subsystems exhibits a rich variety of exotic spin and charge arrangements. As a consequence, it makes both the chains as well as the ladders extremely sensitive to variations of external parameters, such as temperature, doping, and pressure. Therefore, the evolution of the electronic and magnetic structure upon Ca substitution is one of the key issues for the development of a microscopic understanding in such a complex system.

\par

Electron energy-loss spectroscopy (EELS) is a useful tool for the investigation of materials at all levels of complexity in the electronic spectrum~\cite{Fink2001}. The EELS cross section is  proportional to the loss function
$\Im [-1/\epsilon(\omega,\bf q)$], where $\epsilon(\omega,\bf q)$ = $\epsilon_1(\omega,\bf q)$ + $i \epsilon_2(\omega,\bf q)$ is the momentum and energy-dependent complex dielectric function. In this way, EELS probes the collective electronic excitations of a solid  under investigation. Furthermore, it allows momentum dependent measurements of the loss
function, i.\,e., the observation of non-vertical transitions within the band structure of a solid~\cite{Fink1989,Roth2014_EELS}.

\section{Experimental}
\noindent Single crystals of R$_x$Sr$_{14-x}$Cu$_{24}$O$_{41}$ (with R = La and Ca) were grown by using the traveling solvent floating zone
method~\cite{Ammerahl1998}. For the EELS measurements, thin films ($\sim$\,100\,nm) were cut perpendicular to the crystal $b$-axis from these single crystals
using an ultramicrotome equipped with a diamond knife. The films were then put onto standard transmission electron microscopy grids and
transferred into the spectrometer. The measurements as a function of Ca substitution were carried out at room temperature with a dedicated transmission electron energy-loss
spectrometer \cite{Fink1989,Roth2014_EELS} employing a primary electron energy of 172\,keV. The energy and momentum resolution was set to be $\Delta
E$\,=\,80\,meV and $\Delta q$\,=\,0.035\,\AA$^{-1}$, respectively. In addition, the spectrometer is equipped with a He-cryostat, which allows investigating the loss function down to a sample temperature of 20\,K, which has been done for Ca$_{11.5}$Sr$_{2.5}$Cu$_{24}$O$_{41}$. Before measuring the loss-function, the thin films have been characterized by \textit{in-situ} electron diffraction, in order to orient the crystallographic axis with respect to the
transferred momentum~\cite{Roth2010}.

\section{Experimental Results}

\begin{figure}[ht]
\includegraphics[width=0.4\textwidth]{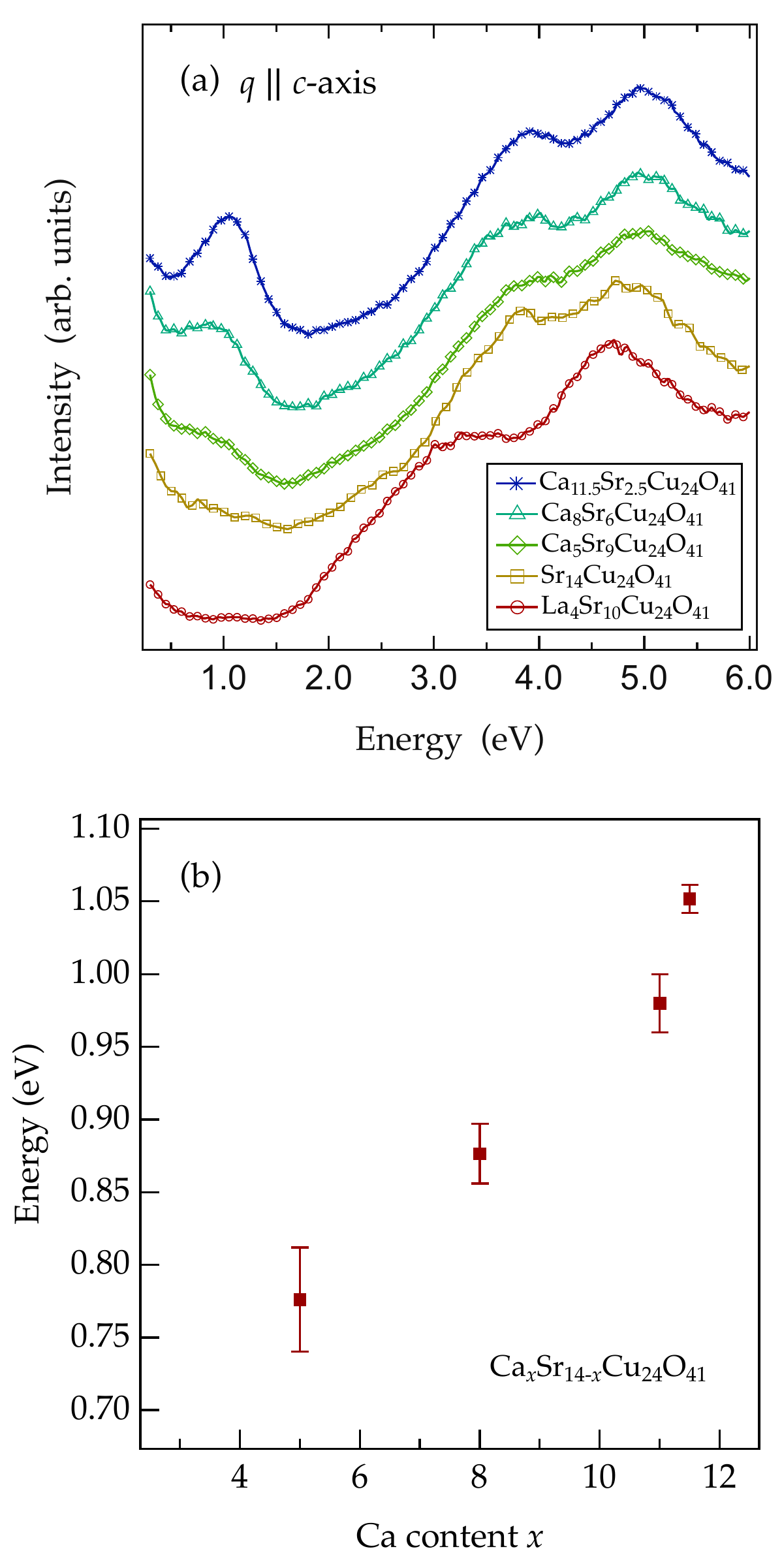}
\caption{\label{fig2} (a) Loss functions for a small momentum transfer of 0.15\,\AA$^{-1}$~parallel to the $c$\,-\,axis for various (La,\,Y,\,Sr,\,Ca)$_{14}$Cu$_{24}$O$_{41}$ compounds. The upturn near zero loss enery is due to the quasi-elastic line. (b) Energy position of the charge carrier plasmon as a function of the Ca concentration. The red squares represent the plasmon position evaluated from the measured spectra for $q$ = 0.15\,\AA$^{-1}$}.
\end{figure}

\noindent We start the presentation of our results with the evolution of the loss function as a function of doping, for a momentum transfer of 0.15\,\AA$^{-1}$~parallel to the $c$\,-\,axis, as shown in Fig.\,\ref{fig2}\,(a).
In the energy range between 3 to 5\,eV we see interband induced collective excitations which have been also detected in the edge-sharing CuO$_2$ chain compounds Li$_2$CuO$_2$ or CuGeO$_3$ ~\cite{Atzkern2000,Atzkern2001,Fink2001}. According to our previous analysis, these excitations are localized excitations in the CuO$_4$ placket.  They are localized because the hopping between the plackets in these one-dimensional edge-sharing Cu-O chains is small. Starting from the undoped compounds the energy of these excitations slightly increases with increasing doping concentration. These excitations are also detected for a momentum transfer parallel to the $a$-axis (see supplementary information and Ref.~\cite{Roth2010}). This observation supports the interpretation in terms of localized excitations. 

\begin{figure}[t]
 \includegraphics[width=0.45\textwidth]{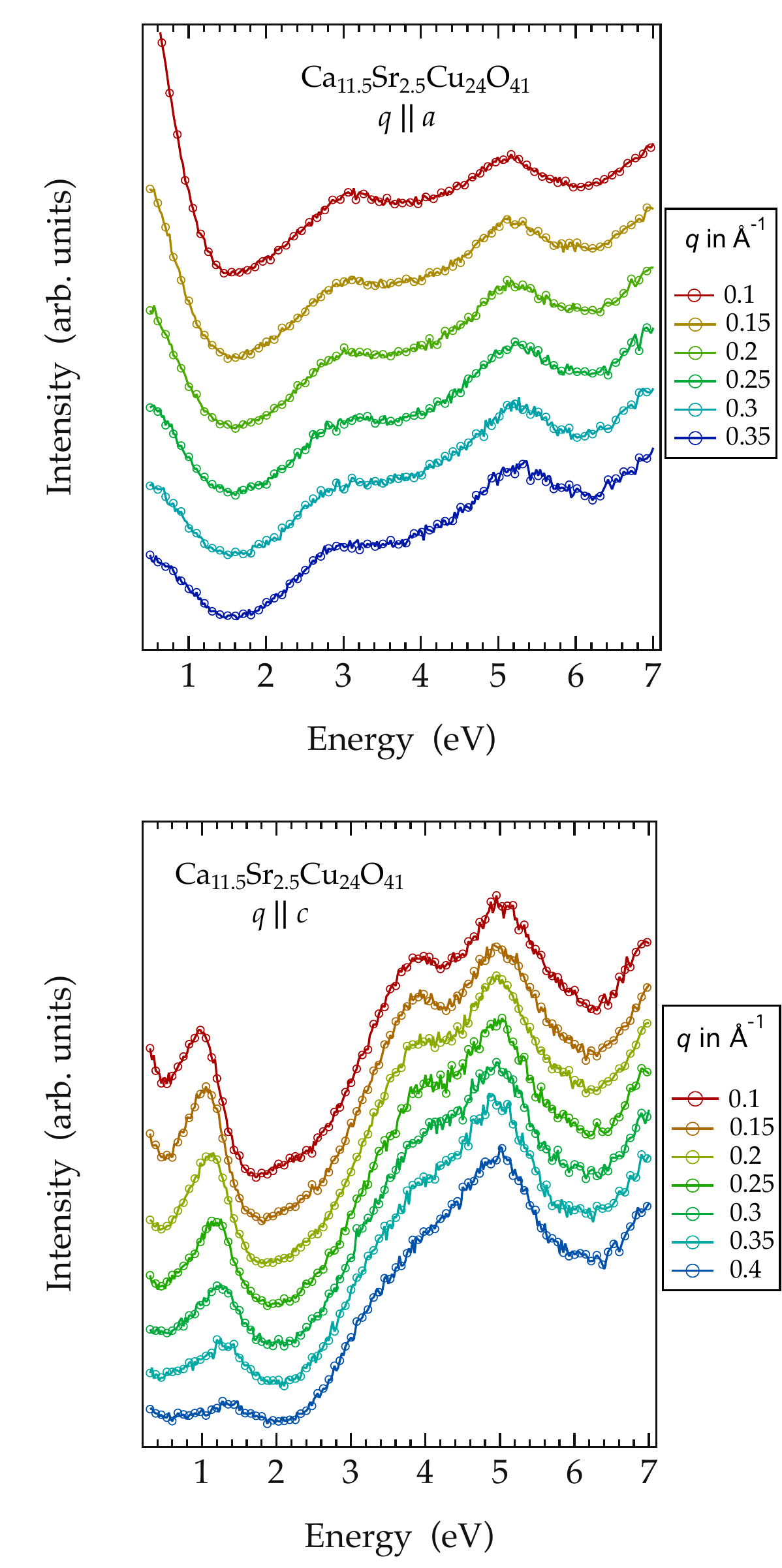}
\caption{\label{fig3} Momentum dependence of the loss function of Ca$_{11.5}$Sr$_{2.5}$Cu$_{24}$O$_{41}$ for a momentum transfer  $q$ parallel to the crystallographic $a$ (upper panel) and $c$ (lower panel) axis. The data have been taken at 300\,K.}
\end{figure}

In the energy  range 2\,-\,3\,eV there are dispersive excitations which according to our previous studies of  the one dimensional corner sharing CuO$_3$ and CuO$_2$ chain compounds Sr$_2$CuO$_3$ and SrCuO$_2$~\cite{Neudert1998,Fink2001}, were ascribed to more delocalized charge transfer excitations~\cite{Wray2008,Wray2008_2,Ishii2007}, also detected in 2D cuprates. These excitations are more pronounced for $q$ parallel to the $a$-axis compared to $q$ parallel to the $c$-axis (see supplementary information and Fig.~2).

\par

 Finally, a plasmon excitation is observed for $x\ge0$ while for the undoped compound the plasmon is absent. Fig.\,\ref{fig2}\,(b) depicts the respective plasmon energy with increasing Ca concentration. The observed plasmon energy is in very good agreement with previous data from reflectivity and EELS measurements (cf. Ref.\,\cite{Osafune1997,Ruzicka1998,Roth2010}). Such a plasmon near 1\,eV was also observed in the two-dimensional doped cuprates~\cite{Nuecker1989}.

\par

In Fig.\,\ref{fig3} we present the loss function for momentum parallel to the $a$ and $c$ direction in Ca$_{11.5}$Sr$_{2.5}$Cu$_{24}$O$_{41}$ at room temperature. Clearly, the plasmon moves to higher energies upon increasing the momentum transfer. The higher energy features, in contrast, are almost independent of momentum. A plasmon dispersion has also been observed for other investigated Ca$_{x}$Sr$_{14-x}$Cu$_{24}$O$_{41}$ ($x$\,=\,5, 8, 11) compounds, and we summarize these results in Fig.\,\ref{fig4} and in the supplementary information. For $q$ parallel $a$ the charge transfer excitation near 3 eV is visible for all Ca concentrations.

\begin{figure}[t]
\includegraphics[width=0.45\textwidth]{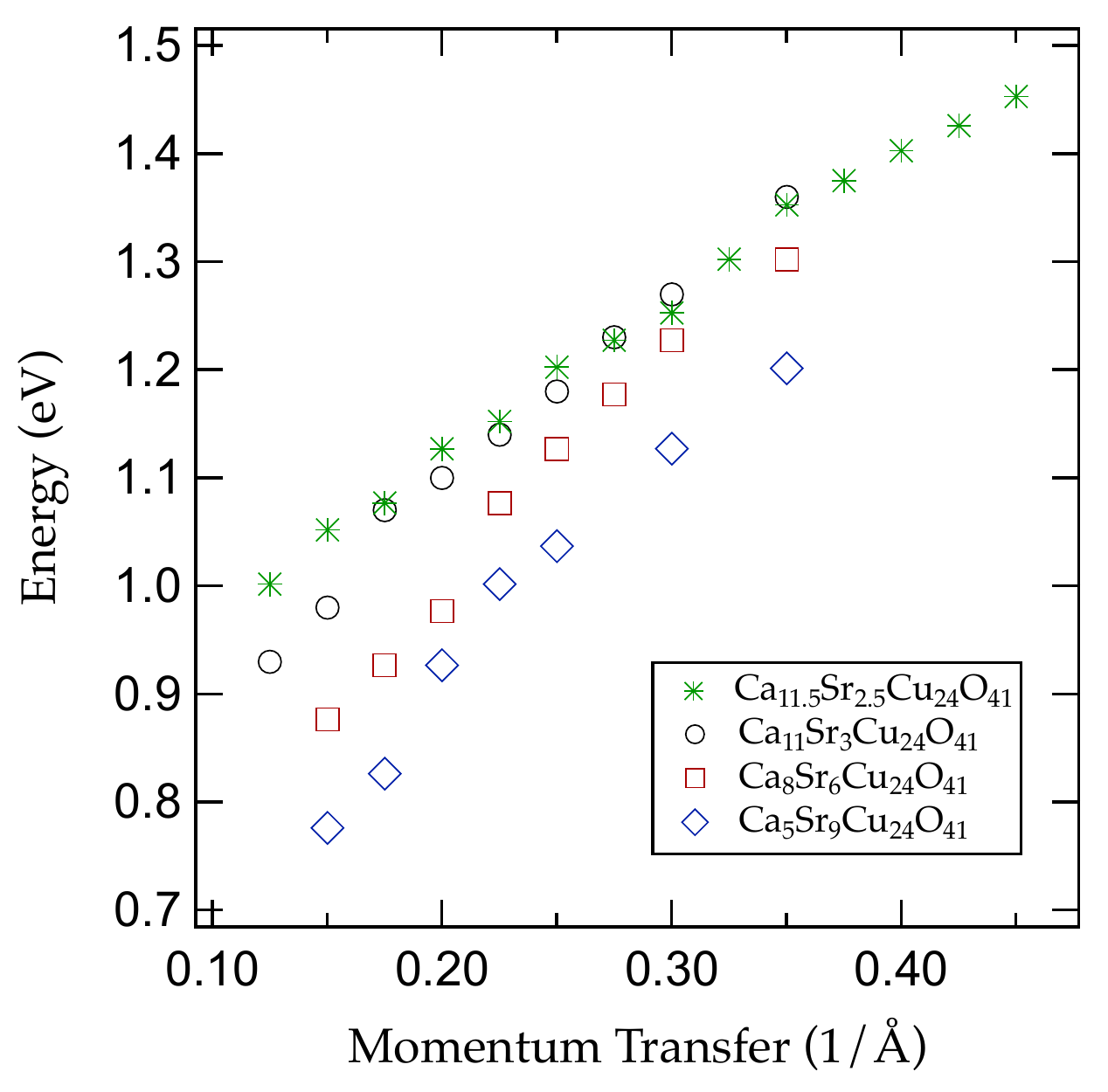}
\caption{\label{fig4}Comparison of the plasmon dispersion in  for various (Ca,\,Sr)$_{14}$Cu$_{24}$O$_{41}$ compounds along the $c$ direction at 300\,K.}
\end{figure}

Fig.\,\ref{fig4} shows again the increasing plasmon energy at low momentum transfers with increasing Ca content in the ladder compounds (cf. Fig.\,\ref{fig2}\,b)). Furthermore, it demonstrates that for all compositions we observe  a finite plasmon intensity, dispersing to higher energies for a momentum transfer parallel to the $c$-axis (notice that for $x$ = 5, 8 and 11 the plasmon peak is only detectable up to a momentum transfer of about 0.35\,\AA$^{-1}$, which is caused by the strong damping as well as the low cross-section for higher momentum transfers). Fig.\,\ref{fig4} reveals another intriguing observation. The slope of the plasmon dispersion also depends on the Ca substitution level. With increasing Ca content the slope of the dispersion curves is reduced. 

\par

\begin{figure}[t]
\includegraphics[width=0.45\textwidth]{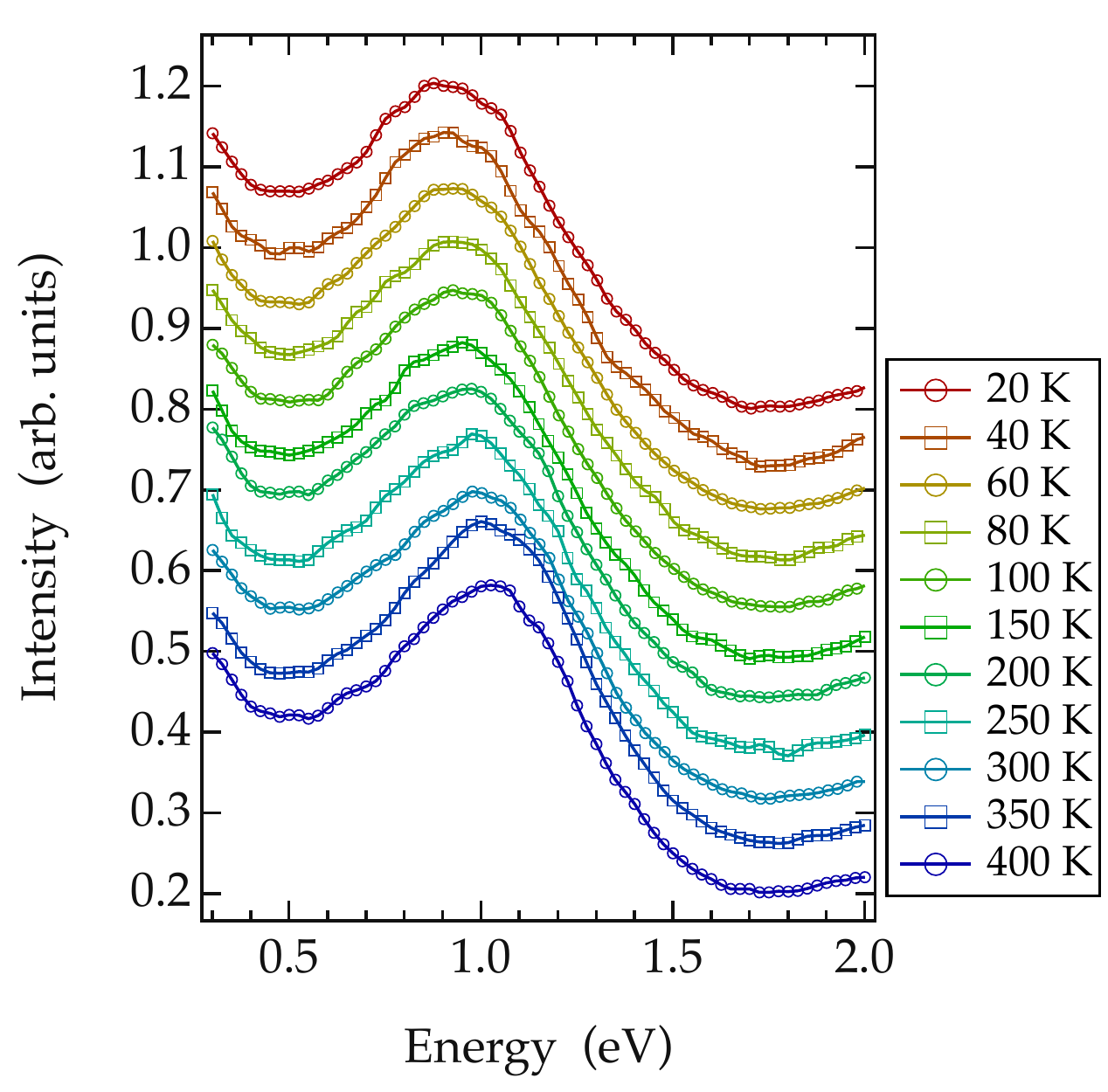}
\caption{\label{fig5}Temperature dependence of the charge carrier plasmon in Ca$_{11.5}$Sr$_{2.5}$Cu$_{24}$O$_{41}$ for a momentum of 0.1\,\AA$^{-1}$ parallel to the c-axis.}
\end{figure}

\begin{figure}[ht]
\includegraphics[width=0.45\textwidth]{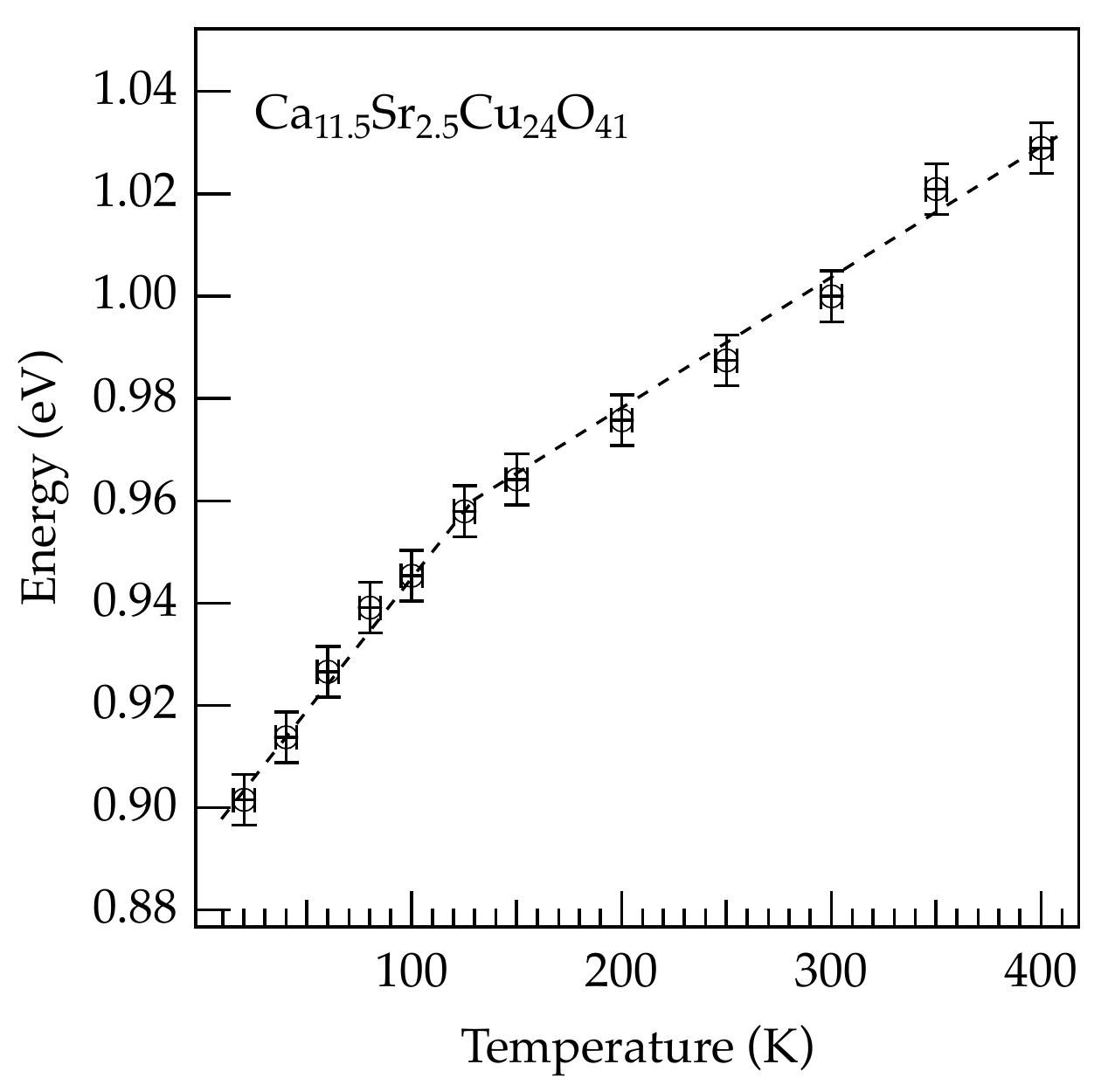}
\caption{\label{fig6}Temperature dependent energy position of the charge carrier plasmon in Ca$_{11.5}$Sr$_{2.5}$Cu$_{24}$O$_{41}$ for a small momentum of 0.1\,\AA$^{-1}$.}
\end{figure}

Finally, we consider the temperature dependence of the plasmon energy in the compound Ca$_{11.5}$Sr$_{2.5}$Cu$_{24}$O$_{41}$. In Fig.\,\ref{fig5} we show the temperature dependence of the plasmon feature in the loss function at a small momentum transfer of 0.1\,\AA$^{-1}$ measured at various temperatures in the range between 20\,K and 400\,K. As shown in Fig.\,\ref{fig6}, the plasmon energy  clearly shifts to higher energies with increasing temperature. Moreover, the data indicate a somewhat stronger shift at lower temperatures (below about 150\,K) as compared to the high-temperature region.

\section{Calculations}

\begin{figure}[ht]
\includegraphics[width=0.49\textwidth]{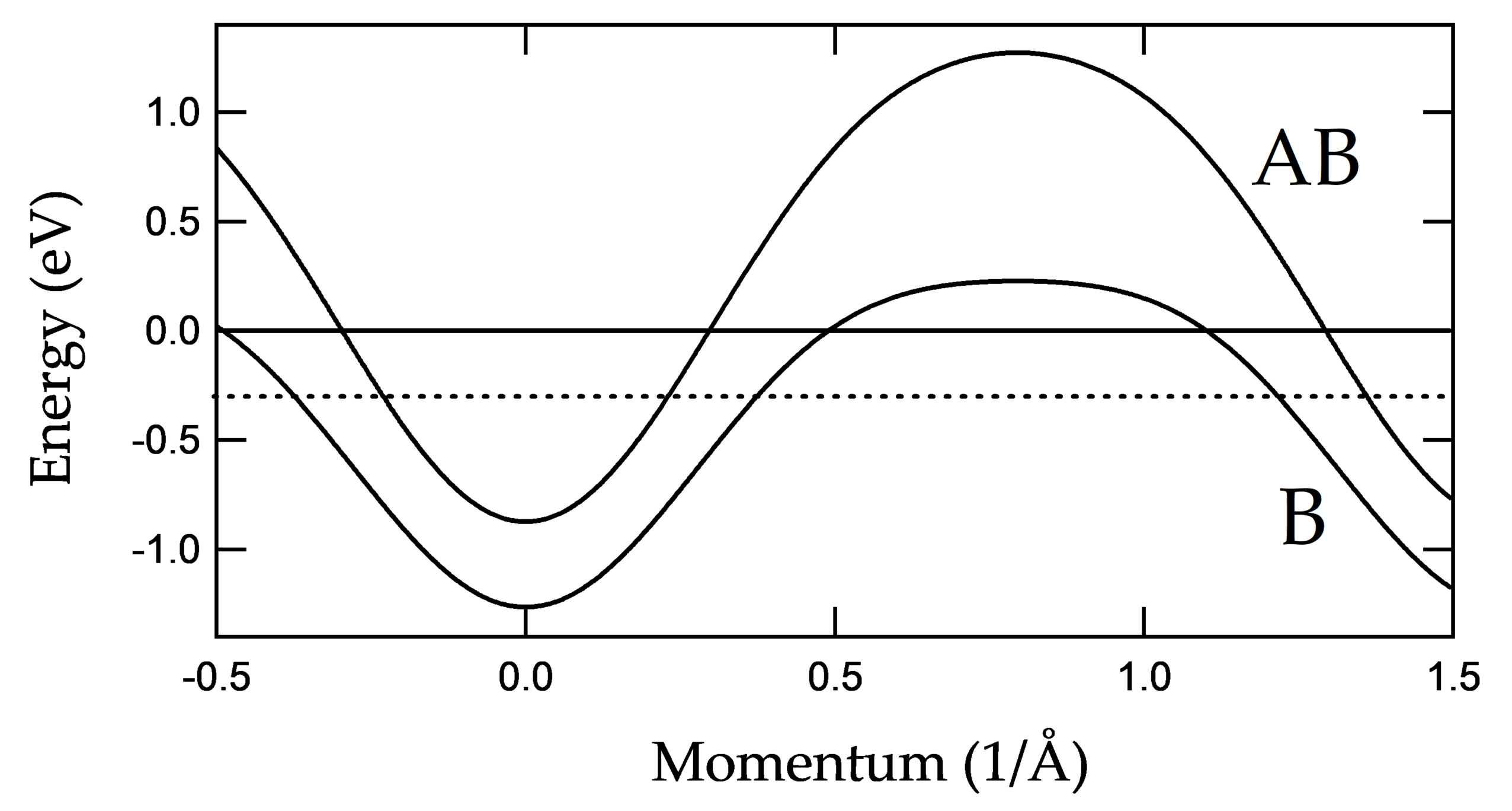}
\caption{\label{fig10} Typical bonding (B) and antibonding (AB) band for the ladders in Ca$_{x}$Sr$_{14-x}$Cu$_{24}$O$_{41} $ along the $c$-axis used in the present calculations.  The solid black horizontal line marks the Fermi level for an almost undoped (relative to divalent Cu) compound while the dashed line marks a Fermi level for a ladder which is about 20 \% doped relative to divalent Cu.
}
\end{figure}

Experiments using optical spectroscopy and EELS, which are related to two-particle excitations, are normally much less sensitive to correlation effects than ARPES in which a single hole is excited. The reason for this is that dipoles are much less screened compared to a single hole. Nevertheless, in optical spectroscopy and EELS of cuprates clear differences in the spectra are observed  between  undoped and  optimally doped compounds. In the former a Mott-Hubbard gap is observed while in the latter, due to a reduction of correlation effects, the gap is closed and  a plasmon is detected~\cite{Uchida1991,Fink2001}. In this less correlated case the plasmon dispersion in optimally doped  Bi$_2$Sr$_2$CaCu$_2$O$_8$ 
could be well described by RPA-type calculations~\cite{Nuecker1991}. Even the anisotropy of the band and the Fermi surface could be confirmed by this combination of EELS derived plasmon dispersions and those calculations. Therefore we started our RPA-type calculations of the plasmon dispersion in
Ca$_{x}$Sr$_{14-x}$Cu$_{24}$O$_{41}$  on the basis of  a tight-binding band structure for the weakly-doped ladders derived from the DFT approximation~\cite{Arai1997}. There are several experimental results that indicate that the edge-sharing chains have a gap larger than 2 eV and therefore do not contribute to the plasmon  excitations in the energy range between 0.5  to 1.5 eV.  In Fig.\,\ref{fig10} we present the band structure for the Cu$_2$O$_3$ ladders calculated using the  tight-banding parameters  of the DFT calculations~\cite{Arai1997}. In this context, we  point out that the band structure of the bonding band detected by ARPES~\cite{Koitzsch2010} is very close to the DFT results.

For the calculation of the complex dielectric function we used the Ehrenreich-Cohen expression~\cite{Ehrenreich1959}:
\begin{equation}\label{EC}
\epsilon (\omega,{q})=\epsilon _{\infty}-\frac{A}{q^2}\int_{-\infty}^{\infty} \frac{\Delta F \Delta E}{\omega^2-\Delta E^2+i\Gamma \omega } dk.
\end{equation}
Here $\Delta E=E_{k+q}-E_k$ and $\Delta F=F_{k+q}-F_k$ where $F$ is the Fermi function. Using the complex dielectric function we then calculate the loss function  $\Im[-1/\epsilon(\omega ,q)]$. Calculating the maximum of the loss function  for a particular momentum $q$ as a function of energy  yields  the plasmon dispersion as a function of the momentum $q$.
The effective mass for the band renormalization was adjusted to the experiment in such a way that the plasmon energy at the highest $q$ values corresponds to the experimental data. The reason for this is that the plasmon energy at high $q$ is essentially determined by the total bandwidth. Mass enhancements between 1.1 and 1.3 have been used for the calculations.  On the other hand, the background dielectric function was adjusted in order to fit the plasmon energy  at low momentum transfer, and it was kept constant thereafter.

\par

\begin{figure}[tb]
\includegraphics[width=0.5\textwidth]{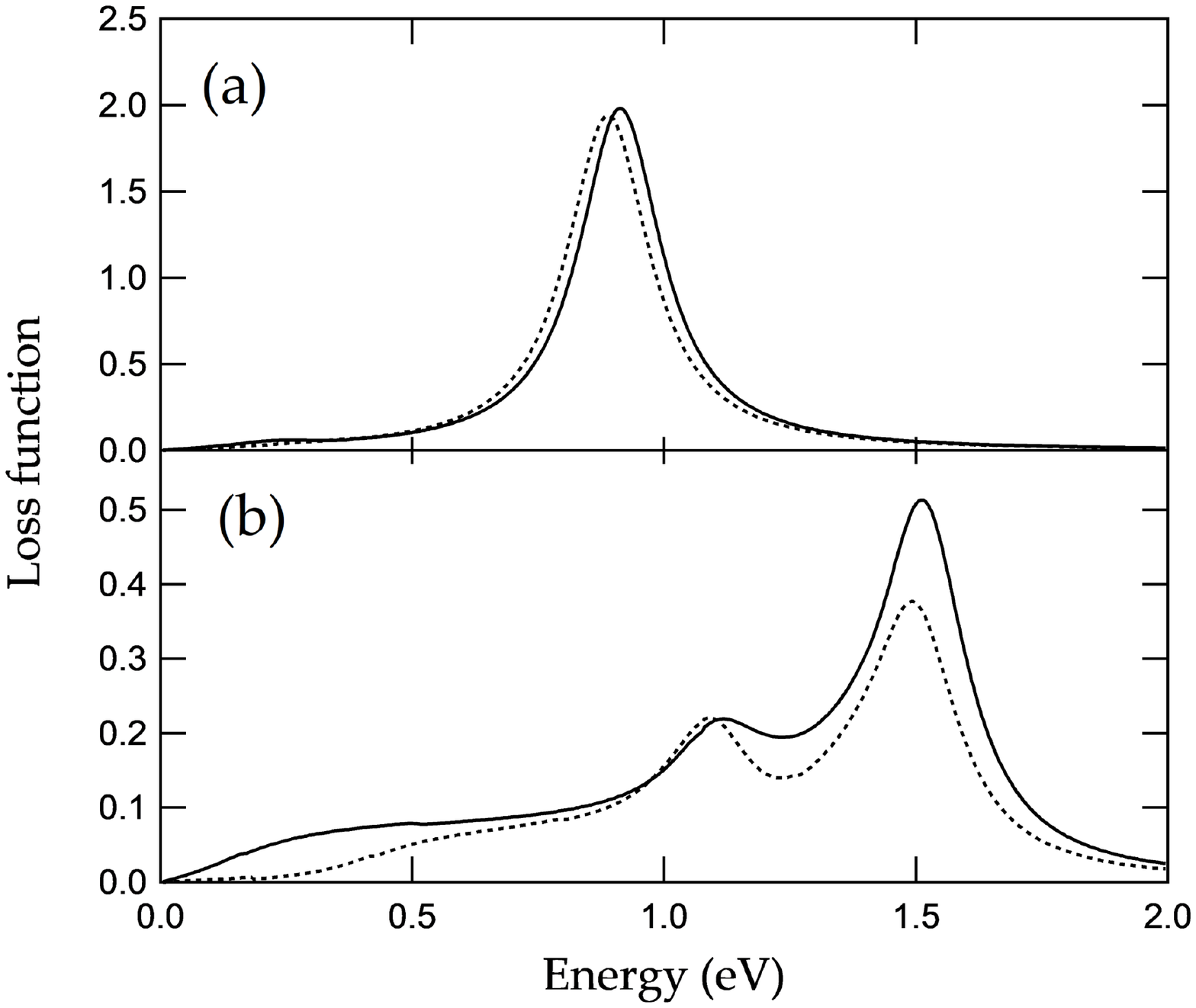}
\vspace{-5.5 cm}
\caption{\label{fig11}
Calculated energy dependence of the loss function $\Im (-1/\epsilon )$ for intra-band  transitions in the anti-bonding and bonding band  in Ca$_{x}$Sr$_{14-x}$Cu$_{24}$O$_{41}$ .
Solid line: undoped ladder ($x\approx 0$). Dashed line: doped ladder with a hole concentration of about 20 \% corresponding to $x\approx14$. (a) Momentum transfer $q$=0.1\,\AA$^{-1}$. (b) Momentum transfer $q$=0.45\,\AA$^{-1}$.
}
\end{figure}

In Fig.\,\ref{fig11} we present the calculated loss function for the sum of intra-band transitions in the bonding and anti-bonding band. Inter-band transitions are not allowed due to symmetry reasons.
There are two maxima in the loss functions corresponding to plasmons related to intra-band transitions in the bonding and the anti-bonding band. At low momentum transfer, the plasmon of the bonding band is strongly over-damped due to the large real part of the dielectric function from the anti-bonding band in that energy range. At higher momentum transfer both plasmons are clearly visible. The fact that two plasmons are observed in the calculations but only one is detected in the EELS spectrum as well as almost no shifts are observed in the calculated spectrum while large shifts are observed upon doping at small momentum transfer
indicates that the starting point in the calculation is wrong.

\par

\begin{figure}[tb]
\includegraphics[width=0.5\textwidth]{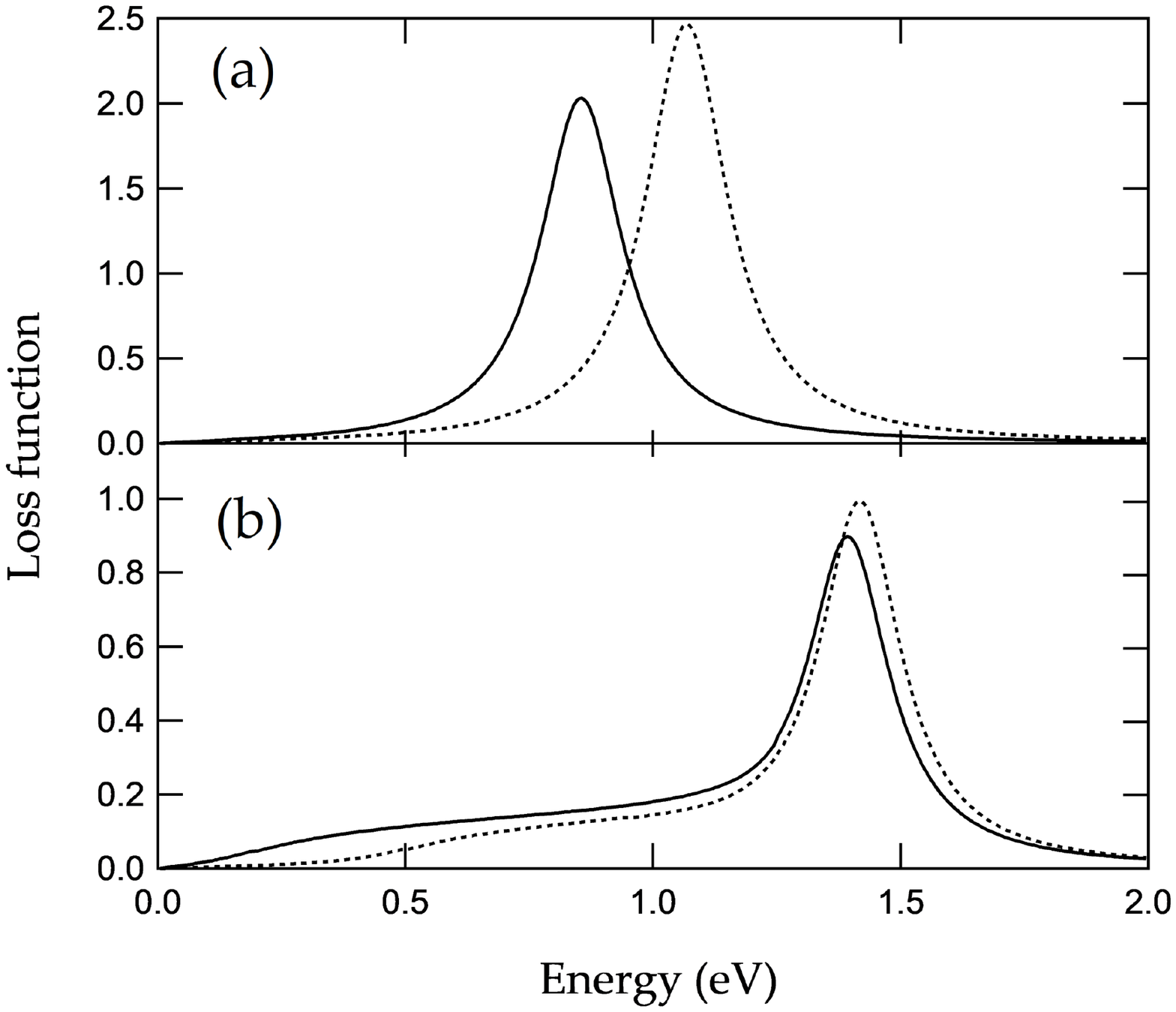}
\vspace{-5.5 cm}
\caption{\label{fig12}
Calculated energy dependence of the loss function $\Im (-1/\epsilon )$) for intra-band  transitions in the  bonding band  in Ca$_{x}$Sr$_{14-x}$Cu$_{24}$O$_{41} $ .
Solid line: undoped ladder ($x\approx 0$). Dashed line: doped ladder with a hole concentration of about 20 \% corresponding to $x\approx14$.  (a) Momentum transfer $q$=0.1\,\AA$^{-1}$. (b) Momentum transfer $q$=0.45\,\AA$^{-1}$.
}
\end{figure}

Much better  agreement between experiment and calculations can be achieved when we assume that only the bonding band contributes to the measured spectral weight of the loss function. The results for this model are shown in  Fig.\,\ref{fig12}. Now we observe a larger shift of the plasmon energy at small momentum transfer $q$=0.1\,\AA$^{-1}$ upon replacement of Sr by Ca  in agreement with the experimental results. At high momentum transfer, the shift is strongly reduced.

\par

\begin{figure}[tb]
\includegraphics[width=0.45\textwidth]{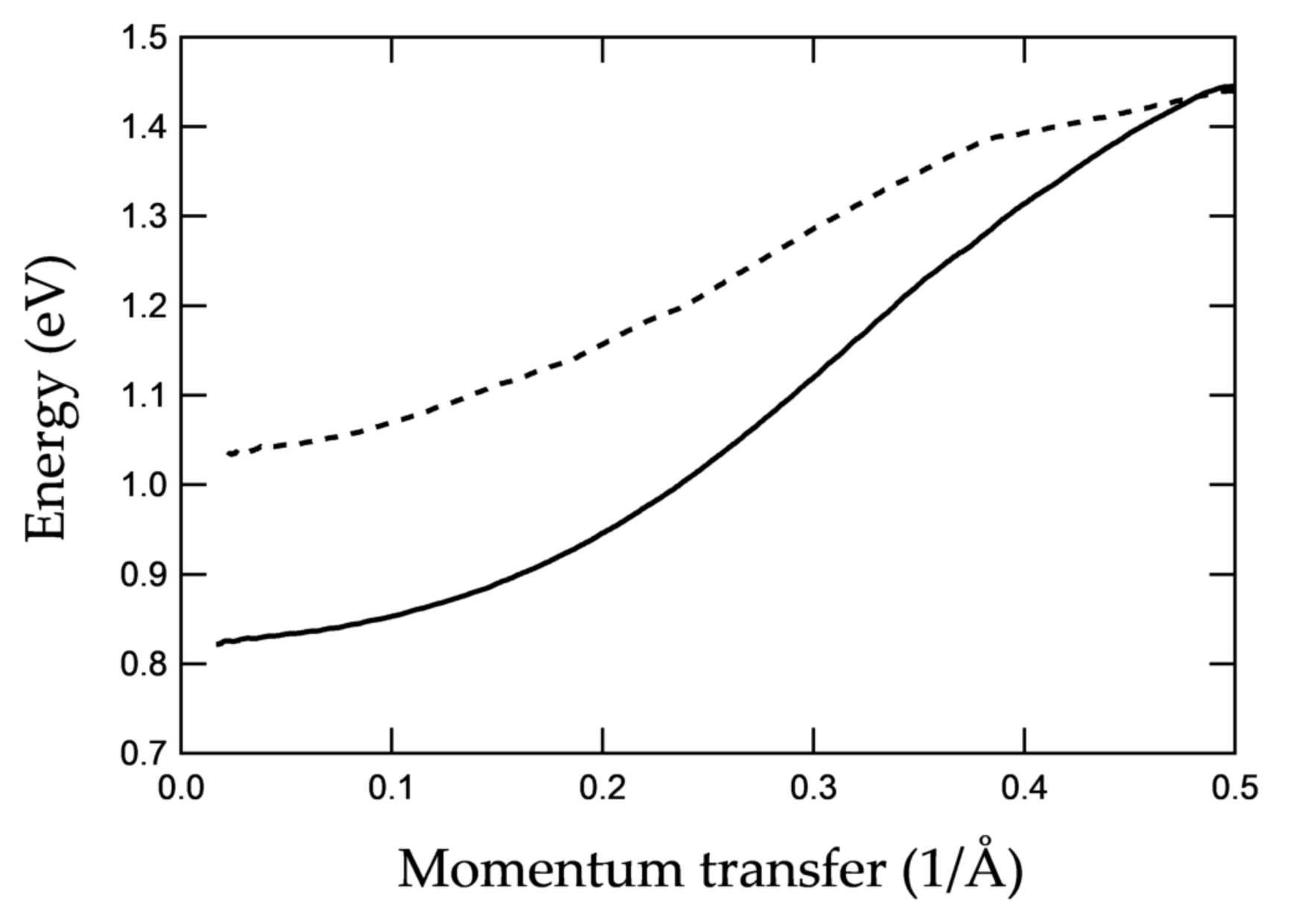}
\caption{\label{fig13}
Calculated plasmon dispersion (momentum dependence of the maximum of the loss function $\Im (-1/\epsilon )$) for intra-band  transitions in the  bonding band  in Ca$_{x}$Sr$_{14-x}$Cu$_{24}$O$_{41} $ .
Solid line: undoped compound ($x\approx 0$). Dashed line: doped compound with a hole concentration of about 20 \% corresponding to $x\approx14$.
}
\end{figure}

In Fig.\,\ref{fig13} we present the  plasmon dispersion  evaluated from the maximum of the plasmon.
Again these results show a large shift of the plasmon energy upon Ca doping at low momentum transfer while at higher $q$ the shift is reduced. The explanation of this result is that at low $q$ the plasmon is determined by the Fermi velocity~\cite{Nuecker1989} which changes considerably upon doping. 
At higher $q$ the plasmon energy is more determined by the bandwidth which does not change upon doping.

\section{Discussion}

Firstly we discuss the higher energy excitations. Looking at  the peaks near 5 eV, they show almost no dispersion for $q||a$ and $c$ (see supplementary information), typical of a localized excitation in edge-shared CuO$_2$ chains,  existing in  Li$_2$CuO$_2$ or CuGeO$_3$ ~\cite{Atzkern2000,Atzkern2001,Fink2001}. Also the peak near 3\,eV shows almost no  dispersion in agreement with previous  EELS studies on edge-shared CuO$_2$  chains.  Compared to the undoped  corner-sharing quasi one-dimensional Sr$_2$CuO$_3$~\cite{Neudert1998} and the double chain SrCuO$_2$ ~\cite{Fink2001}, the charge transfer peak at 2\,-\,3\,eV is shifted into the energy range 3-4\,eV. This peak is also visible at 2.7\,eV in the undoped  2D cuprates, e.\,g., in Sr$_2$CuO$_2$Cl$_2$~\cite{Fink1997,Fink2001}.  The results of resonant inelastic x-ray scattering (RIXS) of Ca$_{x}$Sr$_{14-x}$Cu$_{24}$O$_{41}$ (Ref.\,\cite{Wray2008,Wray2008_2,Ishii2007}) are in  good agreement with our assignment of the spectral structures in the energy range  between 2\,-\,5\,eV.

\par

Next, we discuss the plasmon excitation which is the central point of the present study. Previous studies of Ca$_{x}$Sr$_{14-x}$Cu$_{24}$O$_{41}$ clearly indicated that upon Ca doping there is a transfer of holes from the chain to the ladders~\cite{Osafune1997,Nuecker2000,Isobe2000,Rusydi2007,Deng2011,Ilakovac2012,Huang2013,Bugnet2016}. This indicates that the plasmon is due to excitations in the ladders. The chains do not contribute to the plasmon excitation because the holes are still localized although the doping concentration is rather high. In the unsubstituted compound, the valency of Cu is near  +2.6. The reason that the chains remain gaped and do not contribute to the plasmon excitation  is that the hoping integral is probably much smaller than the onsite Coulomb interaction $U$. This view is in line with the weak reduction of the spectral weight in the upper Hubbard band upon doping~\cite{Nuecker1989}. Thus the plasmon excitation exclusively occurs in the ladders. Most of the previous studies indicate that the number of holes on the ladders of Sr$_{14}$Cu$_{24}$O$_{41}$ is small and of the order of 1 per formula unit, corresponding to a Cu valency of +2.05. The increase of the plasmon energy with increasing Ca substitution [see Fig.1~(b)]  has also been discussed  in a previous optical study~\cite{Osafune1997}.

\par

Our data clearly signal that there is a visible temperature dependence of the hole number in the ladders in Ca$_{11.5}$Sr$_{2.5}$Cu$_{24}$O$_{41}$ (see Figs.\,\ref{fig5} and \ref{fig6}). Evidence for such a hole reduction upon cooling had been reported previously from structural investigations and the related analysis of the bond-valence sum~\cite{Isobe2000,Deng2011}. The observed changes in bond lengths and angles together with the bond-valence sums have been interpreted in terms of a slight hole back-flow to the chains at low temperatures. This scenario is supported by the present EELS data. The small temperature-dependent change of the slope of the plasmon energy below 200\,K (see Fig. \,\ref{fig6}) may be  related to the proximity to a transition into charge density wave (CDW) state in which for smaller $x$ the holes are ordered below
$T_c=250$ K~\cite{Blumberg2002,Vuletic2003,Abbamonte2004}. Possibly the localization of the holes in the CDW state may cause a reduction of the plasmon energy. We emphasize that the opening of a small CDW gap of the order of 0.1\,eV  would only slightly increase the plasmon energy. On the other hand, a mass enhancement in the CDW state would decrease the plasmon energy and could thus explain the shift to lower energies at lower temperatures.

\par

As described in  Section IV the calculation of the loss function for a sum of uncorrelated bonding and anti-bonding band predicts that in this case the loss function is dominated by the plasmon of the anti-bonding band, because that of the bonding band is strongly damped.  The energy of an intra-band plasmon at low momentum transfer is directly related to the Fermi velocity~\cite{Nuecker1991}. This Fermi velocity does almost not change for the anti-bonding band upon doping (see Fig. \,\ref{fig10}). Therefore, the difference in the energy between low and high Ca concentration would be expected to be small.  Shifts of the plasmon related to the bonding band can hardly  be detected because of the low intensity of that plasmon.

\par

On the other hand, the calculation using the bonding band only leads to results that are close to the presented EELS data (see Figs.\,\ref{fig4} and  \ref{fig13}). In this case, the Fermi velocity strongly changes with increasing Ca concentration. For small doping, it is low because the Fermi level is close to the top of the bonding band. At higher hole doping the Fermi level moves to higher binding energies (see Fig. \,\ref{fig10}) and therefore the Fermi energy increases leading to a higher plasmon energy which is in agreement with the experimental EELS data. At higher momentum, the plasmon energy is determined by the full width of the conduction band which only slightly changes upon doping. This is again in agreement with the experiment. From these results we conclude that the nearly half-filled anti-bonding band remains gapped due to correlation effects and that the holes which are transferred from the chains to the ladders upon Ca substitution move into the  bonding band. The correlation effects have a weaker impact on the bonding band because it is only weakly doped. The conclusion that only the bonding band contributes to the plasmon excitation is also supported by the result, that at higher momentum only one plasmon is observed while in the case when both bands are active, the calculation predicts two plasmons (see Fig. \,\ref{fig11}). Finally, the result that a model which does not take into account correlation effects cannot describe the measured loss function is fostered by the consideration that in the uncorrelated case at a small hole concentration ($x=0$) the half-filled anti-bonding band should be related to a strong plasmon excitation. In the experiment, however, for $x=0$ the plasmon is hardly detectable [see Fig.\,\ref{fig2} (a)].

\par

The bonding band, having most of the charge density between the legs of the ladder, has probably mainly $2p$ character from the O ions on the rungs. On the other hand, the anti-bonding band has probably mainly O $2p$  character  on the legs  of the ladders on O $2p$ orbitals which could be parallel $a$ or $c$. Our observation of holes on the O sites in the rungs supports the conception that hole doping of the symmetric bonding band leads to an antisymmetric spin-singlet of the rungs. This means that the low energy charge density wave and superconductivity is based on singlets on the rungs. Our interpretation of the difference between bonding and anti-bonding band is  supported by the observation that even for high Ca concentrations, the charge transfer excitation is detected for $q||a$ (see Fig.~2). This could be related to our postulation that the anti-bonding band is gapped by correlation effects and that the wave function of this anti-bonding band has strong contributions from O $2p$ orbitals parallel to the $a$-axis. We emphasize that from the present work, we only obtain information on the hole distribution between bonding and anti-bonding band. However, we do not obtain any information on the hole distribution on the four O 2p orbitals along the $a$ and the $c$-axis on the rungs and the legs. Possibly calculations beyond the $t-J$ model together with our experimental EELS results could provide more detailed information.

\par

Our interpretation of the plasmon dispersion in these compounds in terms of doping of the rungs is supported by various other experimental results. The XAS results on Ca$_{x}$Sr$_{14-x}$Cu$_{24}$O$_{41}$  indicate that upon replacement of Sr by Ca, the holes move into O $2p$ orbitals on the rungs  while the hole number on O $2p$ orbitals parallel to the legs remain constant~\cite{Nuecker2000}.

In the following, we compare our EELS results with ARPES data of Ca$_{x}$Sr$_{14-x}$Cu$_{24}$O$_{41}$ published in the literature. For the compound without Ca ($x$\,=\,0) Koitzsch et\,al.~\cite{Koitzsch2010} have detected a bonding and an antibonding band, the dispersions of which are close to that derived from DFT calculations~\cite{Arai1997}. It is interesting to note that spectral weight below a binding energy of 0.4\,eV is only detected for the bonding band but not for the anti-bonding. This could be interpreted as a formation of a correlation induced gap for the anti-bonding band but not for the less correlated bonding band. This ARPES result would support our interpretation of the EELS data. On the other hand, for the Ca substituted compound ($x$\,=\,11.5), Koitzsch et\,al. have detected no bonding band, but an anti-bonding band with a clear spectral weight close to the Fermi level. This result would be at variance with our interpretation of the plasmon dispersion which needs a gapped anti-bonding band to explain the absence of a second plasmon. However, in another ARPES study by Yoshida et\,al.~\cite{Yoshida2009} it was emphasized that there is a clear absence of a quasiparticle peak at the Fermi level in the ARPES data of Ca$_{x}$Sr$_{14-x}$Cu$_{24}$O$_{41}$ which is completely different from ARPES spectra of optimally doped high-T$_c$ superconductors with a half-filled band~\cite{Kim2003,Koitzsch2004}. The authors came to the conclusion that there is a gap caused by a charge density wave. Finally, a study by Takahashi et\,al~\cite{Takahashi1997} on the unsubstituted compound ($x$\,=\,0) detected a gap of 0.4\,eV which was assigned to a Mott-Hubbard gap. This would support our analysis of the plasmon dispersion. A further ARPES study by Sato et\,al.~\cite{Sato1998} detected a similar spectral weight even in the substituted sample ($x$\,=\,11.5) indicating a correlation induced gap even in the substituted compound. These findings are in line with the results derived from the analysis of the present plasmon dispersion. Summarizing the comparison presented in this paragraph yields in some cases a support of our EELS analysis by ARPES, in other cases, there exists no agreement. Moreover, we want to emphasize that EELS, in contrast to APRES, is a more bulk-sensitive method---independent from the surface conditions as well as cleanness of the sample. Therefore, the presented results help to get a deeper insight into the microscopic electronic structure of that complex system and can be used to finally clear up the published conflicting ARPES results as discussed above.

\par

The cuprate ladder compounds are in some way the intermediate systems between the doped 1D  systems forming a Luttinger liquid with spinon and holons and the two-dimensional cuprates, which in the overdoped case form a  Fermi liquid. The $t-J$ model~\cite{Zhang1988} is believed to represent the gross features of the electronic structure of the two-dimensional cuprates. In this model, the charge carriers can be described  by singlet  particles formed by  a Cu~$3d$ hole on the divalent Cu site  and a O~$2p$  hole  on the surrounding O square. In theoretical studies, it also was tried to describe the charge carriers in a two-leg ladder by a $t-J$ model~\cite{Dagotto1996,Kumar2019}.  The charge excitations in the Cu-O ladders cannot be described by a one-particle Hamiltonian based on Zhang-Rice singlets. Rather the results point to an orbital dependent Mott transition for the two-leg layer, leading to the formation of mobile singlet holes on the rungs, related to the bonding band and  in the anti-bonding band a Mott-Hubbard state, or more precise a charge transfer insulator. Our results thus may question also the application of the Zhang-Rice singlet model for the precise description of the electronic structure of the two-dimensional cuprate high-$T_c$ superconductors.

\par

\section{Summary}

We have investigated the charge carrier plasmon in the spin-ladder compound Ca$_{x}$Sr$_{14-x}$Cu$_{24}$O$_{41}$  as a function of Ca substitution and temperature using electron energy-loss spectroscopy. The energy of the plasmon increases upon increasing Ca content, which signals an increasing number of holes in the Cu$_2$O$_3$ ladders due to an internal charge redistribution between chains and ladders.  A comparison of the experimental plasmon  excitations with  RPA-like  calculations for the  bonding and anti-bonding band indicates that the holes, which are transferred from the chain and the ladder upon replacement of Sr by Ca, are mainly located in the rungs. The reason for this is the different filling of the bonding and antibonding band leading to a Mott Hubbard splitting for the half-filled anti-bonding band and rather mobile charge carriers in the bonding band having predominantly  O $2p$  character with orbitals along the rungs. Thus the charge carriers cannot be described by  Zhang-Rice singlets very often  employed  for the 2D cuprates. Rather in the ladder system which is between the 1D and the 2D cuprates, the charge carriers are singlets on the rungs fundamentally different from those of the 1D and 2D cuprates. Our results may question the general application of 
 the $t-J$ model to cuprate high-$T_c$ superconductors. 

\begin{acknowledgments}
\noindent We thank M. Naumann, R. H\"ubel, F. Thunig and S. Leger for technical assistance. We acknowledge fruitful discussions with C. Wohlleben,  J. Lorenzano, and R. Eder.

\end{acknowledgments}

 \bibliographystyle{apsrev4-1}

\end{document}